\documentstyle[pra,aps,twocolumn,epsf,amssymb]{revtex}

\begin{document}
\draft
\title{Interferometric Bell-state preparation using femtosecond-pulse-pumped
Spontaneous Parametric Down-Conversion}
\author{Yoon-Ho Kim,\thanks{Email: yokim@umbc.edu} Maria V.
Chekhova,\thanks{Permanent address: Department of Physics, Moscow
State University, Moscow, 119899, Russia.} Sergei P.
Kulik,$^\dagger$ Morton H. Rubin, and Yanhua Shih}
\address{Department of Physics, University of Maryland, Baltimore
County. Baltimore, Maryland 21250}
\date{To appear in Phys. Rev. A (2001)}

\maketitle

\vspace*{-1cm}

\widetext

\begin{abstract}
We present theoretical and experimental study of preparing
maximally entangled two-photon polarization states, or Bell
states, using femtosecond pulse pumped spontaneous parametric
down-conversion (SPDC). First, we show how the inherent
distinguishability in femtosecond pulse pumped type-II SPDC can
be removed by using an interferometric technique without spectral
and amplitude post-selection. We then analyze the recently
introduced Bell state preparation scheme using type-I SPDC.
Theoretically, both methods offer the same results, however,
type-I SPDC provides experimentally superior methods of preparing
Bell states in femtosecond pulse pumped SPDC. Such a pulsed
source of highly entangled photon pairs is useful in quantum
communications, quantum cryptography, quantum teleportation, etc.
\end{abstract}


\narrowtext

\section{Introduction}\label{intro}

The nature of quantum entanglement attracted a great deal of
attention even in the early days of quantum mechanics, yet it
remained an unsolvable subject of philosophy until Bell showed the
possibility of practical experimental tests
\cite{Schrodinger,EPR,Bohm,Bell}. Since then, many experimental
tests on the foundations of quantum mechanics have been performed
\cite{clauser,aspect,shihalley,usingspdc}. All these tests
confirmed quantum mechanical predictions. More recently,
experimental and theoretical efforts are being shifted to
``applications", such as quantum communications, quantum
cryptography \cite{crypto}, and quantum teleportation
\cite{teleportation}, taking advantage of the peculiar physical
properties of quantum entanglement. It is clear that preparation
of maximally entangled two-particle (two-photon) entangled
states, or Bell states, is an important subject in modern
experimental quantum optics.

By far the most efficient source of obtaining two-particle entanglement is spontaneous
parametric down-conversion (SPDC). SPDC is a nonlinear optical process in which a
higher-energy pump photon is converted into two lower-energy daughter photons, usually
called the signal and the idler, inside a non-centrosymmetric crystal \cite{SPDC}. In
type-I SPDC, both daughter photons have the same polarizations but in type-II SPDC, the
signal and the idler photons have orthogonal polarizations. The signal and the idler are
generated into a non-factorizable entangled state. The photon pair is explicitly
correlated   in energy and momentum or equivalently in space and time. To prepare a
maximally entangled two-photon polarization state, or a Bell state, one has to make
appropriate local operations on the SPDC photon pairs. \newpage \vspace*{38mm}

The polarization Bell states, for photons, can be written as
\begin{eqnarray}
|\Phi^\pm\rangle&=&|X_1,X_2\rangle\pm|Y_1,Y_2\rangle,\nonumber\\
|\Psi^\pm\rangle&=&|X_1,Y_2\rangle\pm|Y_1,X_2\rangle,
\label{bellstates}
\end{eqnarray}
where the subscripts 1 and 2 refer to two different photons,
photon 1 and photon 2, respectively, and they can be arbitrarily
far apart from each other. $|X\rangle$ and $|Y\rangle$ form the
orthogonal basis for the polarization states of a photon, for
example, it can be horizontal ( $|H\rangle$) and vertical
($|V\rangle$) polarization state, as well as $|45^\circ\rangle$
and $|-45^\circ\rangle$, respectively.

The subject of this paper is a detailed theoretical and
experimental account of how one can prepare a polarization Bell
state using femtosecond pulse pumped SPDC. In section
\ref{intro}, we discuss why one needs such a pulsed source of Bell
states, what happens when femtosecond pulsed laser is used to
pump a type-II SPDC, and what has been done to recover the
visibility in femtosecond pulse pumped type-II SPDC. In section
\ref{type2theory}, we present a detailed theoretical description
of how one can prepare a polarization Bell state using
femtosecond pulse pumped type-II SPDC without any post-selection,
followed by the experiment in section \ref{type2exp}. We then
turn our attention to type-I SPDC and investigate it in detail
theoretically in section \ref{type1theory} and experimentally in
section \ref{type1exp}.

The quantum nature of SPDC was first studied by Klyshko in late
1960's \cite{klyshkoSPDC}. Zel'dovich and Klyshko predicted the
strong quantum correlation between the photon pairs in SPDC
\cite{zeldovich}, which was first experimentally observed by
Burnham and Weinberg \cite{burnham}. The nonclassical properties
of SPDC were first applied to develop an optical brightness
standard \cite{klyshko2} and absolute measurement of detector
quantum efficiency \cite{calibration}.

Quantum interference effect in SPDC was first clearly demonstrated
by Hong, Ou, and Mandel \cite{HOM}. Shih and Alley first used
SPDC to prepare a Bell state \cite{shihalley}. Such experiments
have used type-I non-collinear SPDC and a beamsplitter to
superpose the signal-idler amplitudes. Experimentally, type-I
non-collinear SPDC is not an attractive way of preparing a Bell
state mostly due to the difficulties involved in alignment of the
system. Collinear type-II SPDC developed by Shih and Sergienko
resolved this issue \cite{shihsergienko}. There is, however, a
common problem: the entangled photon pairs have 50\% chances of
leaving at the same output ports of the beamsplitter. Therefore,
the state prepared after the beamsplitter may not be considered
as a Bell state without amplitude post-selection as pointed out
by De Caro and Garuccio \cite{garuccio}. Only when one considers
the coincidence contributing terms by throwing away two out of
four amplitudes (post-selection of 50\% of the amplitudes), the
state is then considered to be a Bell state. Kwiat \emph{et al}
solved this problem by using non-collinear type-II SPDC
\cite{kwiat1}. This non-collinear type-II SPDC method of
preparing a Bell state has been widely used in quantum optics
community.

Recently, cw pumped type-I SPDC has also been used to prepare Bell
states. Kwiat \emph{et al} used two thin nonlinear crystals to
prepare Bell states using non-collinear type-I SPDC \cite{kwiat2}
and Burlakov \emph{et al} used a beamsplitter to join collinear
type-I SPDC from two thick crystals in a Mach-Zehnder
interferometer type setup \cite{Burlakov}.

Therefore, in cw pumped SPDC, there are readily available
well-developed methods of preparing a Bell state. However,
entangled photon pairs occur randomly within the coherence length
of the pump laser beam. This huge time uncertainty makes it
difficult to use in some applications, such as generation of
multi-photon entangled state, quantum teleportation, etc, as
interactions between entangled photon pairs generated from
different sources are required. This difficulty was thought to be
solved by using a femtosecond pulse laser as a pump.
Unfortunately, femtosecond pulse pumped type-II SPDC shows poor
quantum interference visibility due to the very different
(compared to the cw case) behavior of the two-photon effective
wave-function \cite{keller}. One has to utilize special
experimental schemes to maximize the overlap of the two-photon
amplitudes. Traditionally, the following methods were used to
restore the quantum interference visibility in femtosecond pulse
pumped type-II SPDC: (i) to use a thin nonlinear crystal ($\leq
100 \mu$m) \cite{sergienko} or (ii) to use narrow-band spectral
filters in front of detectors \cite{femto}. Both methods,
however, reduce the available flux of the entangled photon pair
significantly and cannot achieve complete overlap of the
wave-functions in principle. We will discuss this in detail in
section \ref{type2theory}.

Branning \emph{et al} first reported an interferometric technique to remove the
intrinsic distinguishability in femtosecond pulse pumped type-II SPDC without using
narrowband filters and a thin crystal \cite{branning}. This method, however, cannot be
used to prepare a Bell state since \emph{four} biphoton amplitudes are involved in the
quantum interference process \cite{notebranning}. This issue will be discussed in section
\ref{type2theory}. More recently, Atat\"{u}re \textit{et al.} claimed recovery of
high-visibility quantum interference in pulse pumped type-II SPDC from a thick crystal
without spectral post-selection. Unfortunately, the theory as well as the interpretation
of the experimental data presented their work are shown to be in error \cite{comment}.
Then it is fair to say that there have been no generally accepted method of preparing a
Bell state from femtosecond pulse pumped SPDC without making any post-selection,
especially the spectral post-selection. In the following sections, we will present
experimental studies of preparing a Bell state using femtosecond pulse pumped SPDC (both
type-II and type-I) together with the theoretical analysis.

\section{Bell state preparation using type-II SPDC: Theory}\label{type2theory}

\begin{figure}[b]
\centerline{\epsfxsize=3.2in\epsffile{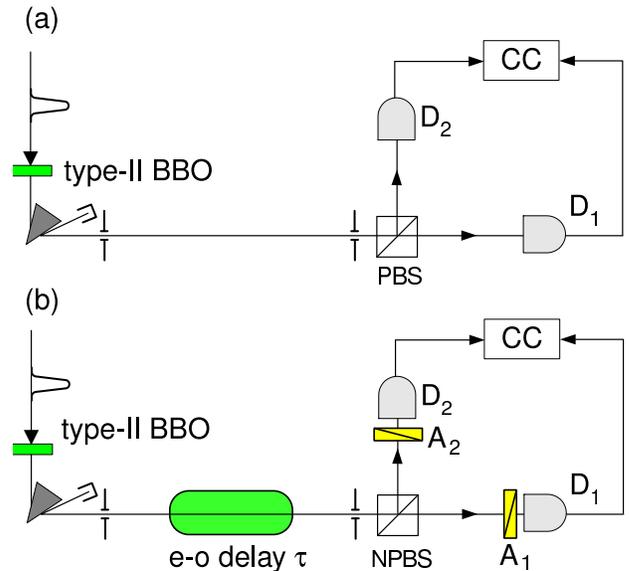}} \caption{(a) Most simple
two-photon correlation experiment. A femtosecond pulse pumps a type-II BBO crystal to
create entangled photon pairs via SPDC process. Orthogonally polarized signal and idler
photons are separated by a polarizing beam splitter ($PBS$) and detected by two
detectors $D_1$ and $D_2$. (b) Collinear Hong-Ou-Mandel interferometer to observe
quantum interference. The e-o delay $\tau$ can be introduced by a stack of quartz
plates. $NPBS$ is a 50/50 non-polarizing beam splitter. $A_1$ and $A_2$ are polarization
analyzers. }\label{type2spdc}
\end{figure}

Let us first briefly discuss the basic formalism of pulse pumped
type-II SPDC as discussed by Keller and Rubin \cite{keller}. A
femtosecond pulse pumps a type-II BBO crystal to create entangled
photon pairs via SPDC process. Orthogonally polarized signal and
idler photons are separated by a polarizing beam splitter (PBS)
and detected by two detectors, see Fig.\ref{type2spdc}(a).

Let us start from the Hamiltonian of the SPDC \cite{keller,Rubin}
\begin{equation}
{\mathcal H}=\epsilon_0\int
d^3\vec{r}\chi^{(2)}E_p(z,t)E_o^{(-)}E_e^{(-)},
\end{equation}
where $E_p(z,t)$ is the electric field for the pump pulse which
is considered to be classical, and $E_o^{(-)}$ ($E_e^{(-)}$) is
the negative frequency part of the quantized electric field for
the $o$-polarized ($e$-polarized) photon inside the $\chi^{(2)}$
nonlinear crystal (BBO). The pump field can be written as
\begin{equation}
E_p(z,t)={\mathcal E}_p\int d\omega_p e^{-[\omega_p-\Omega_p]^2/\sigma_p^2}
e^{i[k_p(\omega_p)z-\omega_p t]},
\end{equation}
where ${\mathcal E}_p$ is the amplitude of the pump pulse, $\Omega_p$ is the central
frequency of the pump pulse, $\sigma_p^2=4\ln2/[\sigma^{\rm FWHM}_p]^2$ where
$\sigma^{\rm FWHM}_p$ is the FWHM bandwidth of the pump pulse, and $z$-direction is taken
to be the pump pulse propagation direction. In the interaction picture, the state of
SPDC is calculated from first-order perturbation theory \cite{Rubin}
\begin{eqnarray}
|\psi\rangle&=&-\frac{i}{\hbar}\int_{-\infty}^{\infty}dt
{\mathcal H}|0\rangle, \nonumber \\
&=&C \int dk_o dk_e d\omega_p \int_o^L dz e^{-[\omega_p-\Omega_p]^2/\sigma_p^2}e^{i
\Delta z} \nonumber \\ & & \times \delta(\omega_o+\omega_e-\omega_p) a_o^\dagger
a_e^\dagger |0\rangle, \label{psi}
\end{eqnarray}
where $C$ is a constant, $L$ is the thickness of the crystal,
$a_o^\dagger$ ($a_e^\dagger$) is the creation operator of
$o$-polarized ($e$-polarized) photon in a given mode, and
$\Delta=k_p-k_o-k_e$ the phase mismatch.

The state vector $|\psi\rangle$ obtained in Eq.(\ref{psi}) is used
to calculate the probability of getting a coincidence count
\cite{Glauber}.
\begin{equation}
R_c\propto \int dt_1 \int dt_2
|\langle0|E_2^{(+)}E_1^{(+)}|\psi\rangle|^2,
\end{equation}
where the field at the detector $D_1$ can be written as
\begin{equation}
E_1^{(+)}=\int d\omega' e^{-[\omega'-\Omega_1]^2/\sigma_1^2 }
e^{-i\omega't_1^o}a_o(\omega'),\label{field}
\end{equation}
where $t_1^o=t_1-l_1^o/c$, $l_1^o$ is the optical path length experienced by the
$o$-polarized photon from the output face of the crystal to $D_1$ and $a_o(\omega')$ is
the destruction operator of $o$-polarized photon of frequency $\omega'$. $\Omega_1$ is
the central frequency and $\sigma_1^2=4\ln2/[\sigma_1^{\rm FWHM}]^2$ where $\sigma^{\rm
FWHM}_1$ is the FWHM bandwidth of the spectral filter inserted in front of the detector
$D_1$. $E_2^{(+)}$ is defined similarly.

We now define the two-photon amplitude (or biphoton) as
\begin{equation}
A(t_+,t_-)=\langle0|E_2^{(+)}E_1^{(+)}|\psi\rangle,
\end{equation}
where $t_+ \equiv (t_1^o+t_2^e)/2$, and $t_- \equiv t_1^o-t_2^e$.

For generality, we have included the spectral filtering in
Eq.(\ref{field}). The effect of spectral filtering on the
two-photon effective wave-function in femtosecond pulse pumped
type-II SPDC is studied theoretically and experimentally by Kim
\emph{et al} \cite{kim1}. For the purpose of this paper, the
bandwidths of the spectral filters $\sigma_1$ and $\sigma_2$ are
now taken to be infinite. Let us also assume degenerate SPDC
($\Omega_1=\Omega_2$).

Therefore the two-photon amplitude originated from each pump
pulse has the form \cite{keller}
\begin{eqnarray}
A(t_+,t_-)&=&e^{-i\Omega_p t_+} \int_{-\infty}^\infty
d\nu_p\int_{-\infty}^\infty d\nu_-\int_0^{L}dz\nonumber\\
&\times&e^{-[\nu_p/\sigma_p]^2} e^{-i\left\{\nu_p D_+ +
[\nu_-/2]D\right\}z} \nonumber \\ &\times& e^{-i\nu_p t_+}e^{-i[\nu_-/2]t_-}, \nonumber\\
&\equiv&e^{-i\Omega_pt_+}\Pi(t_+,t_-),
\end{eqnarray}
where $D_+ \equiv \frac{1}{2}
\left\{1/u_o(\Omega_o)+1/u_e(\Omega_e)\right\} -1/u_p(\Omega_p)$,
and $D \equiv 1/u_o(\Omega_o)-1/u_e(\Omega_e)$. $u_o(\Omega_o)$
is the group velocity of o-polarized photon of frequency
$\Omega_o$ inside the BBO. Subscripts $o$, $e$, and $p$ refer to
o-polarized photon, e-polarized photon, and the pump,
respectively. $\nu_p$ is the detuning from the pump central
frequency $\Omega_p$ ($\nu_p=\omega_p-\Omega_p$). $\nu_o$ and
$\nu_e$ are defined similarly and $\nu_-\equiv\nu_o-\nu_e$.

The exact form of the $\Pi(t_+,t_-)$ function is given by
$$
\Pi(t_+,t_-) =
    \left\{
        \begin{array}{ll}
        g e^{-\sigma_p^2\{t_+ - [D_+/D] t_-\}^2/4} & {\rm for~} 0<t_-<DL \\
        0 & {\rm otherwise}
        \end{array}
    \right.
$$
where $g$ is a constant. Note that, different from the cw case
where $\Pi$ is a function of $t_-$ only \cite{Rubin}, $\Pi$ is
now a function of both $t_+$ and $t_-$.

\begin{figure}[tbp]
\centerline{\epsfxsize=3.4in\epsffile{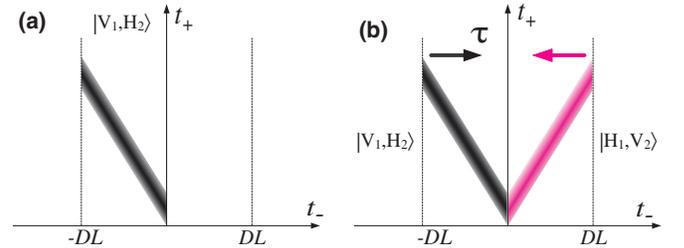}} \caption{(a) Biphoton
wave-function with femtosecond pulse pump in the case of Fig.\ref{type2spdc}(a) shown in
density plot. It differs significantly from the case of cw pumped type-II SPDC. It
starts at $-DL$ and ends at $0$ due to the fact that a BBO is a negative uniaxial
crystal so that the group velocity of the e-polarization is greater than that of the
o-polarization. This figure shows how the two-photon amplitude $|V_1,H_2\rangle$ is
distributed in time. (b) Two biphoton amplitudes ($|V_1,H_2\rangle$ and
$|H_1,V_2\rangle$) are present in the case of the experimental setup shown in
Fig.\ref{type2spdc}(b). When $\tau$ is increased, the two biphoton wave-functions move
toward each other. If the pump is cw, the biphoton is essentially infinitely long in
$t_+$ direction. If $\tau=DL/2$, in this case, the overlap is complete and 100\% quantum
interference can be observed. However, there is only limited amount of overlap due to
the peculiar shape of the biphoton amplitudes in pulse pumped type-II SPDC and this
results in the reduction of the visibility of quantum interference. This is the inherent
difference from cw pumped type-II SPDC.}\label{type2biphoton}
\end{figure}

The shape of $\Pi(t_+,t_-)$ function is shown in
Fig.\ref{type2biphoton}(a). It differs from the cw pumped type-II
SPDC significantly. Similar to the cw case, the biphoton starts
at $t_-=0$ and ends at $|t_-|=DL$ \cite{Rubin}, but unlike the cw
case, there is a strong dependence on $t_+$ direction. This is
the reason why quantum interference visibility is reduced in
femtosecond pulse pumped type-II SPDC.

To prepare a polarization entangled state using type-II collinear SPDC, one first has to
replace the polarization beam splitter ($PBS$) in Fig.\ref{type2spdc}(a) with a
non-polarizing 50/50 beam splitter ($NPBS$), see Fig.\ref{type2spdc}(b), so that there
are two biphoton amplitudes to contribute to a coincidence count: (i) the signal is
transmitted and the idler is reflected at the $NPBS$ ($t-r$ or $|V_1,H_2\rangle$
amplitude), or (ii) the signal is reflected and the idler is transmitted at the $NPBS$
($r-t$ or $|H_1,V_2\rangle$ amplitude). Here we only considered the coincidence
contributing amplitudes: amplitude post-selection. When these two $t-r$
($|V_1,H_2\rangle$) and $r-t$ ($|H_1,V_2\rangle$) amplitudes are made indistinguishable,
a Bell state is prepared (modulo amplitude post-selection) and it can be confirmed
experimentally by observing 100\% quantum interference \cite{shihsergienko}. To make the
two amplitudes indistinguishable, the e-o delay $\tau$ should be correctly chosen. A
typical method to find the correct e-o delay $\tau$ is to observe Hong-Ou-Mandel dip
when the e-o delay $\tau$ in Fig.\ref{type2spdc}(b) is varied. One then fixes $\tau$
where the complete destructive (when analyzers are set at $A_1=A_2=45^\circ$) or
constructive (when analyzers are set at $A_1=45^\circ$ and $A_2=-45^\circ$) interference
occurs.

What happens in the experimental setup shown in
Fig.\ref{type2spdc}(b) can be understood easily in the biphoton
picture shown in Fig.\ref{type2biphoton}(b). As discussed above,
there are two biphoton amplitudes distributed in ($t_+$,$t_-$)
space. The one on the left represents $|V_1,H_2\rangle$ and the
one on the right represents $|H_1,V_2\rangle$. When the e-o delay
$\tau=0$, there is no overlap, i.e, no quantum interference. As
$\tau$ increases, the biphoton wave-functions move toward each
other by $\tau$. In cw pumped type-II SPDC, when $\tau=DL/2$, the
overlap between two amplitudes is complete since the biphoton
amplitude is essentially independent of $t_+$, i.e., 100\%
quantum interference can be observed. On the other hand, in
femtosecond pulse pumped type-II SPDC, as shown in
Fig.\ref{type2biphoton}(b), the amount of overlap is very small
even at $\tau=DL/2$. Due to the tilted shape of the biphoton
amplitude, there can never be 100\% overlap between the two
amplitudes and this results in the reduction of the visibility of
quantum interference. It is important to note that, by
introducing $\tau$, we are shifting the biphoton amplitudes in
$t_-$ direction only.

There are several ways to increase the overlap between the two
biphoton amplitudes: (i) One can use a thin BBO crystal. In this
case the relative area of overlap between the two biphoton
amplitudes is increased (since $|DL|$ is decreased) by sacrificing
the amount of photon flux. (ii) One can use very narrowband
spectral filters in front of the detectors. In this case, the
biphoton amplitudes get broadened strongly in $t_-$ direction,
which results in increased overlap between the two amplitudes
(The effect of spectral filtering in $t_+$ direction is much
smaller than that  in $t_-$ direction) \cite{kim1}. Again, the
available photon flux is reduced.

\begin{figure}[tbp]
\centerline{\epsfxsize=2in\epsffile{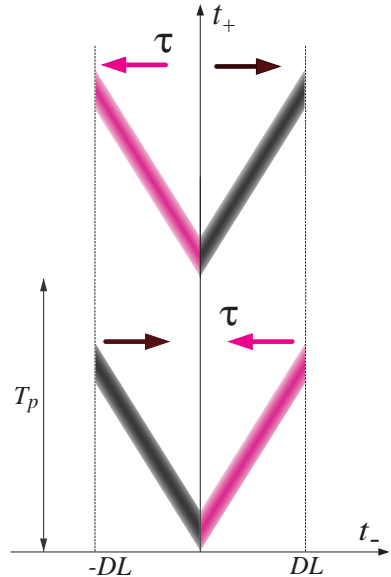}} \caption{Biphoton amplitudes for
Branning \emph{et al}'s scheme. Four amplitudes are involved in the quantum
interference, two from the first pass of the pump pulse (upper diagram) and two from the
second pass of the pump pulse (lower diagram). The delay between the amplitudes from the
first-pass and the second-pass is $T_p$. $\tau$ is the same e-o delay shown in
Fig.\ref{type2spdc}(b). The direction of arrows represent how the relevant amplitudes
moves in $t_-$ direction when $\tau$ is increased. }\label{branning}
\end{figure}

Branning \emph{et al} recently introduced an interferometric technique to overcome this
problem by placing a type-II BBO crystal in a Michelson interferometer \cite{branning}.
Such a method can in principle give a 100\% quantum interference. It, however, cannot be
used to prepare a polarization Bell state since there are \emph{four} biphoton
amplitudes, rather than two, involved in the interfering process \cite{notebranning}. In
addition, the first-order interference (observed in single counting rates) cannot be
avoided in Branning \emph{et al}'s scheme \cite{origin}. Let us discuss this a little
further. By placing a thick (5mm) type-II BBO crystal into a Michelson interferometer,
Branning \emph{et al} achieve a double-pass down-conversion scheme \cite{notebranning2}.
In this case, there are \emph{four} biphoton amplitudes involved in the process: two from
the first-pass of the pump pulse and the other two from the second-pass of the pump pulse
since each pass of the pump pulse results in two biphoton amplitudes as shown in
Fig.\ref{type2biphoton}(b). Then the non-zero contribution of the biphoton amplitudes in
Branning \emph{et al}'s scheme can be depicted as in Fig.\ref{branning}. $T_p$ is the
delay introduced between the first-pass and the second-pass of the pump pulse, i.e. the
delay between the two biphoton amplitudes from the first-pass of the pump pulse and the
two biphoton amplitudes from the second-pass of the pump pulse. This delay $T_p$ is only
introduced in $t_+$ direction. The e-o delay $\tau$ is introduced in $t_-$ direction by
introducing a stack of quartz plates as before, see Fig.\ref{type2spdc}(b) and
Fig.\ref{type2biphoton}(b). When $\tau=0$ and $T_p=0$, 100\% quantum interference should
be observed if polarization information is erased by setting both analyzers at
$45^\circ$. However, Bell states of the type shown in Eq.(\ref{bellstates}) have not
been prepared in this method.

\begin{figure}[tbp]
\centerline{\epsfxsize=3.2in\epsffile{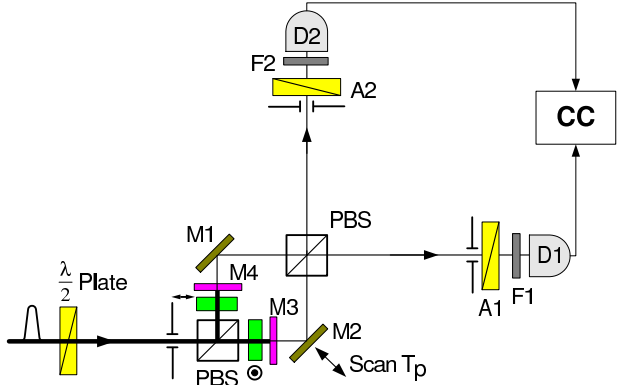}} \caption{Bell state preparation
scheme using two type-II BBO in a Mach-Zehnder interferometer. $T_p$ is the delay
between the biphoton amplitudes from two different crystals. Therefore, the relative
phase between the two amplitudes, $\varphi = \Omega_p T_p$. Note that the optic axes of
the BBO crystals are oriented orthogonally, one vertically ($\updownarrow$) and the other
horizontally ($\odot$). This is important with femtosecond pump pulse, but with cw pump
it is of no importance. See text for detail. }\label{type2setup}
\end{figure}

Let us now consider the experimental setup shown in
Fig.\ref{type2setup} \cite{similar}. A type-II BBO crystal is
placed in each arm of the Mach-Zehnder interferometer (MZI). The
pump pulse is polarized at $45^\circ$ by using a $\lambda/2$
plate. PBS is the polarizing beam splitter. The optic axis of the
first BBO is oriented vertically ($\updownarrow$) and the other
horizontally ($\odot$). The pump pulse is blocked by the mirrors
$M_3$ and $M_4$. There are only two biphoton amplitudes in this
process, one from each crystal, due to the fact that a polarizing
beam splitter is used to split the signal and the idler of the
entangled photon pairs. Therefore, the quantum state, when the
MZI is properly aligned, can be written as
\begin{equation}
|\psi\rangle= |H_1,V_2\rangle + e^{i\varphi}
|V_1,H_2\rangle,\label{MZItypeII}
\end{equation}
where $H$ and $V$ represent horizontal and vertical polarization respectively.
Therefore, by varying the relative phase delay $\varphi$, one can prepare the Bell
states $|\Psi^{\pm}\rangle$. The other two Bell states $|\Phi^{\pm}\rangle$ can be easily
achieved by inserting a $\lambda/2$ plate in one output port of the MZI. Note that
$\varphi=\Omega_p T_p$, where $T_p$ is the time delay between the two amplitudes in
Eq.(\ref{MZItypeII}), so that by varying $T_p$, modulation in the coincidence counting
rate is observed at the pump central frequency. Therefore, in this scheme, a true Bell
state can be prepared without any post-selection methods. As we shall show later in this
section, the thickness of the nonlinear crystals and the spectral filtering of the
entangled photon flux do not affect the visibility in principle, even with a femtosecond
pulse pump.

\begin{figure}[tbp]
\centerline{\epsfxsize=3.2in\epsffile{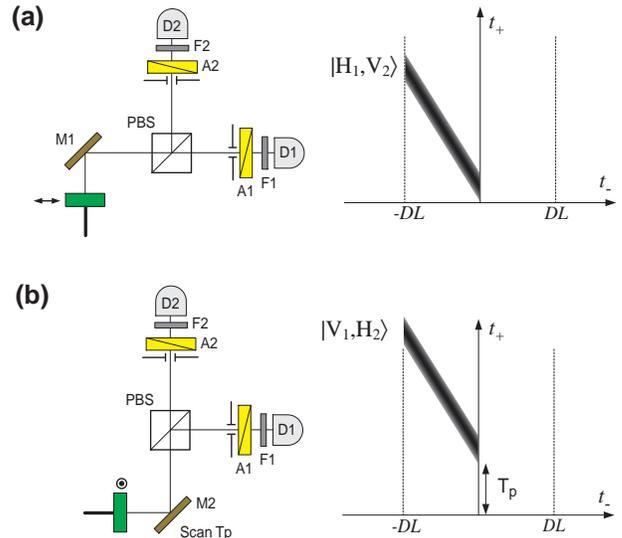}} \caption{This figure
illustrates the two interfering biphoton amplitudes in the experimental setup shown in
Fig.\ref{type2setup}. $T_p$ is the time delay between the two biphoton amplitudes. If
the MZI is balanced, i.e., $T_p=0$, complete overlap can be
achieved.}\label{type2amplitudes}
\end{figure}

In this configuration, e-polarized photons are always detected by
$D_2$ and o-polarized photons are always detected by $D_1$. This
is of great importance when femtosecond laser is used as a pump.
The biphoton amplitude for each coincidence detection event is
shown in Fig.\ref{type2amplitudes}. Note that only two biphoton
amplitudes are involved in the quantum interference. If the MZI
is balanced, 100\% quantum interference can be observed. This
provides a good method of preparing Bell states. If cw pump is
used instead \cite{kwiat3}, it is not absolutely necessary to
have the optic axes of the crystals orthogonally oriented.
Suppose that the optic axis of the crystal in
Fig.\ref{type2amplitudes}(b) is now oriented vertically
($\updownarrow$), then the corresponding biphoton amplitude will
appear flipped about $t_-=0$, thus appearing from 0 to $DL$.
Clearly, there cannot be any overlap between two amplitudes even
with the balanced MZI ($T_p=0$), just as the case considered in
Fig.\ref{type2spdc}(b) and Fig.\ref{type2biphoton}(b). However,
in cw pump case, the biphoton amplitudes are independent of
$t_+$. Therefore, by making appropriate compensation in $t_-$
direction, the two amplitudes can be overlapped: i.e., by setting
the e-o delay $\tau=DL/2$. Note that if the coherence length of
the cw pump laser is not long enough, then perfect overlap cannot
be obtained for the same reason as in femtosecond pulse pumped
case.

So far, we have pictorially shown that a true Bell state can be
obtained in femtosecond pulse pumped type-II SPDC by using an
interferometric technique. The picture we have presented is based
on the biphoton amplitude calculated earlier in this section. The
space-time and polarization interference effects are calculated
as follows. We use the right-hand coordinate system assuming the
direction of propagation as $z$-axis. Then the sum of biphoton
amplitudes in the experimental scheme Fig.\ref{type2setup} is,
\begin{eqnarray}
{\mathcal{A}}(t_+,T_p,t_-)= (\hat{e}_1\cdot\hat{e}_H)(\hat{e}_2\cdot\hat{e}_V)
e^{-i\Omega_p t_+}\Pi(t_+,t_-) +\nonumber \\
(\hat{e}_1\cdot\hat{e}_V)(\hat{e}_2\cdot\hat{e}_H) e^{-i\Omega_p (t_+ + T_p)}\Pi(t_+ +
T_p,t_-),\label{typeIIamplitude}
\end{eqnarray}
where $\hat{e}$ represents the unit vector in a certain direction,
for example, $\hat{e}_1$ represents the direction of the analyzer
$A_1$. The coincidence counting rate is then calculated as
\begin{eqnarray}
R_c&=&\int dt_+ dt_- |{\mathcal{A}}(t_+,T_p,t_-)|^2 \nonumber \\
   &=&\int dt_+ dt_-|\sin\theta_1\cos\theta_2\Pi(t_+,t_-)+\nonumber\\&&\hspace*{0.7in}
   \cos\theta_1\sin\theta_2
   e^{-i\Omega_p T_p}\Pi(t_+ + T_p,t_-)|^2 \nonumber \\
   &\propto& \sin^2\theta_1\cos^2\theta_2+\cos^2\theta_1\sin^2\theta_2+\nonumber \\
   &&2V(T_p)\cos(\Omega_p T_p)
   \sin\theta_1\cos\theta_2\cos\theta_1\sin\theta_2,
\end{eqnarray}
where
\begin{eqnarray}
V(T_p)&\equiv& \frac{\int dt_+ dt_- \Pi(t_+,t_-)\Pi(t_+ + T_p,t_-)}{\int dt_+ dt_-
\Pi^2(t_+,t_-)} \nonumber\\&=& e^{-[\sigma_p T_p]^2/8}.\label{type2envelope}
\end{eqnarray}

Therefore, the space-time interference at
$\theta_1=\theta_2=45^\circ$ will show
\begin{equation}
R_c=1+V(T_p)\cos(\Omega_p T_p),\label{spacetime}
\end{equation}
and the polarization interference will show
\begin{equation}
R_c=\sin^2(\theta_1+\theta_2) \hspace*{0.5cm} {\rm for}
\hspace*{0.5cm} \Omega_pT_p=0.
\end{equation}

It is important to note that the envelope of the space-time
interference when no spectral filters are used, $V(T_p)$
function, is exactly equal to that of the self-convolution of the
pump pulse. No crystal parameters affect the envelope of the
space-time interference pattern. As we shall show in section
\ref{type1theory}, this is a special feature of type-II SPDC. If
any spectral filtering is used, naturally, the envelope will be
broadened.

\section{Bell state preparation using type-II SPDC: Experiment}\label{type2exp}

As mentioned before, the goal of this section is to experimentally demonstrate that
high-visibility quantum interference, which can be used to prepare a two-photon
polarization Bell state, can be observed in the experimental scheme shown in
Fig.\ref{type2setup} and the envelope of this interference fringes is exactly the same
as the pump pulse envelope.

Let us first discuss the experimental setup, see
Fig.\ref{type2setup}. As briefly discussed in section
\ref{type2theory}, a type-II BBO crystal is placed in each arm of
the MZI and the optic axes of the crystals are oriented
orthogonally, one vertically ($\updownarrow$) and the other
horizontally ($\odot$). The thickness of the crystals is 3.4mm
each. The crystals are pumped by frequency-doubled (by using a
700 $\mu$m type-I BBO) radiation of Ti:Sa laser oscillating at
90MHz. The pump has the central wavelength of 400nm. The average
power of the laser beam in each arm of the MZI is approximately
10mW. The residual pump laser beam is blocked by two mirrors
$M_3$ and $M_4$ and the relative phase $\varphi=\Omega_p T_p$
between the two amplitudes can be varied by adjusting one of the
mirrors $M_2$. Collinear degenerate down-conversion is selected
by a set of pinholes.

\begin{figure}[tbp]
\centerline{\epsfxsize=3.2in\epsffile{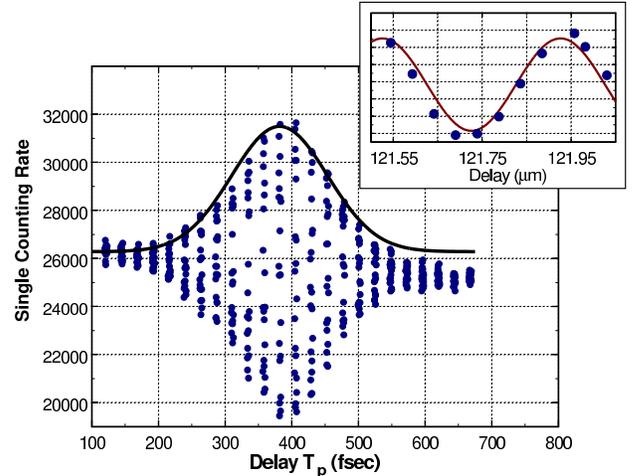}} \caption{Observed pump pulse
interference by blocking the SPDC photons. Solid line is a Gaussian fit to the envelope.
The FWHM of the pump pulse envelope measured from the Gaussian fit is 170fsec. Each
column of data represents the modulation of about one wavelength and the modulation
period is 400nm as shown in the inset. The inset has the same vertical scale as the main
figure and the delay is displayed in $\mu$m (rather than in fsec) to clearly show the
modulation period. The inset shows the detailed modulation around $T_p \approx$ 406
fsec. }\label{pumpenvelope}
\end{figure}

We first measure the envelope of the pump pulse itself by blocking
the SPDC photons while detecting a small fraction of the pump
light that passed through the mirrors $M_3$ and $M_4$. This is
done by simply using another set of interference filters that
transmit 400nm radiation. Fig.\ref{pumpenvelope} shows the
measured envelope of the pump pulse interference patters. The
measured FWHM is 170fsec and this will be compared with the
envelope of the two-photon quantum interference patterns.

The space-time quantum interference is observed at
$\theta_1=\theta_2=45^\circ$ by varying $T_p$. Two sets of
experimental data are collected by using two different sets of
interference filters, FWHM bandwidths of 10nm and 40nm with 800nm
central wavelength, to demonstrate the effect of spectral
filtering on the biphoton wave-function. 3nsec coincidence window
is used and single counting rates of the detectors are recorded
as well.

\begin{figure}[tbp]
\centerline{\epsfxsize=3.2in\epsffile{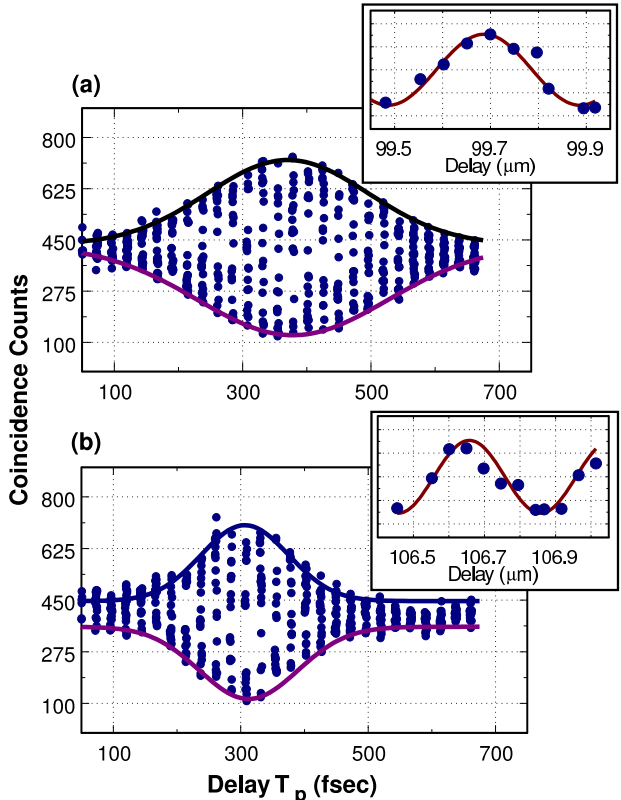}} \caption{Quantum interference
observed with the experimental setup shown in Fig.\ref{type2setup}. (a) 10nm filters are
used. Accumulation time is 60sec. The measured width of the envelope is approximately
310 fsec. The visibility is $\approx$ 75\%. (b) 40nm filters are used. Accumulation time
is 80sec. The measured FWHM of the envelope is $\approx$ 170fsec which is equal to the
FWHM envelope of the pump pulse itself. The visibility is $\approx$ 76\%. Each column of
data represent the modulation of about one wavelength with modulation period of 400nm as
predicted in Eq.(\ref{spacetime}). The insets have the same vertical scales as the main
figures and they show the detailed modulations around (a) $T_p \approx$ 332 fsec and (b)
$T_p \approx$ 356 fsec. }\label{type2-10nm40nm}
\end{figure}

Fig.\ref{type2-10nm40nm} shows the data for these two measurements. In
Fig.\ref{type2-10nm40nm}(a), the FWHM of the interference envelope is 310 fsec. This
shows that 10nm filter has some effect on the shape of the biphoton wave-function. This
is not so surprising since the FWHM bandwidth of the SPDC spectrum for 3.4mm BBO is 3nm.
However, when 40nm filters are used, see Fig.\ref{type2-10nm40nm}(b), the FWHM of the
envelope (170 fsec) is equal to the FWHM of the pump pulse interference patterns, see
Fig.\ref{pumpenvelope}. The effects of the spectral filters are not present and
Eq.(\ref{spacetime}) is confirmed experimentally. The average visibility is 76\% which
is higher than in any femtosecond-pulse pumped type-II SPDC experiments with a thick
crystal and no spectral filters. No interference is observed in the single detector
counting rates.

The visibility loss is mostly due to the imperfect alignment of
the system. Due to the anisotropy of the BBO crystal, e-ray walks
off from the beam path of o-ray. Although both e-rays from the two
different crystals are collected at the same detector, the
walk-off is in different directions: one walks off horizontally,
the other walks off vertically. When thick crystals are used,
3.4mm in our case, such effect is not negligible \cite{walkoff}.
Since we are not interested in making any filtering, spectral or
spatial, spatial filtering using a single mode fiber is not
desirable. Instead of using spatial filtering, such a walk-off
can be removed in another way: the two crystals are oriented in
the same direction and then insert a $\lambda/2$ plate after one
of the crystals \cite{kwiat3}.

The interferometry using the MZI, however, has one disadvantage:
keeping the phase coherence between the two arms of the
interferometer over a long time can be difficult. Although we have
shown here that the visibility can be improved and in principle
reach 100\%, if the phase coherence is not kept for a long time,
such a method is not useful as a source of Bell states for other
experiments. We now turn our attention to type-I SPDC and
investigate whether it offers a good solution to this problem. As
we shall show, type-I SPDC offers a better way of preparing Bell
states in femtosecond pulse pumped SPDC.

\section{Bell state preparation using type-I SPDC: Theory}\label{type1theory}

In this section, we discuss how one can prepare a polarization
Bell state in an interferometric way using \emph{degenerate}
type-I SPDC pumped by a femtosecond pulse pump.

In general, the difference between type-II SPDC and type-I SPDC
stems from calculating the phase mismatch term $\Delta$ in
Eq.(\ref{psi}). In type-II SPDC, due to the fact that the signal
and the idler photons have orthogonal polarization, only the
first-order Taylor expansion of $\Delta$ is necessary, even in
degenerate case. In degenerate type-I SPDC, however, one has to go
to the second-order expansion since the first-order terms cancel
if the frequencies are degenerate. Therefore, non-degenerate
type-I SPDC formalism is basically the same as that of degenerate
type-II SPDC and we will not discuss it here again. On the other
hand, as we shall show, degenerate type-I SPDC differs quite a
lot from type-II SPDC or non-degenerate type-I SPDC.

\begin{figure}[tbp]
\centerline{\epsfxsize=3.2in\epsffile{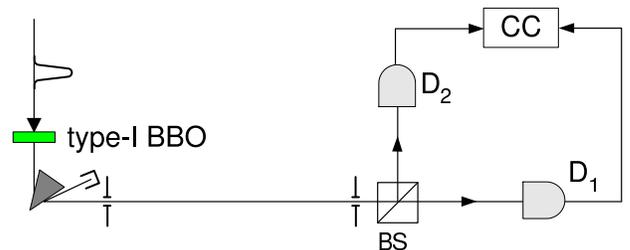}} \caption{Most simple two-photon
correlation experiment using type-I SPDC. See text for detail.}\label{type1spdc}
\end{figure}

Let us first consider the experimental situation shown in
Fig.\ref{type1spdc}. Type-I SPDC occurs in the crystal and the
photon pairs are detected by two detectors $D_1$ and $D_2$. The
state of type-I SPDC is the same as Eq.(\ref{psi}) except that
both photons are now $o$-polarized. For type-I degenerate SPDC,
the phase mismatch term $\Delta$ becomes
\begin{equation}
\Delta = - \left( \nu_p D_+ + \frac{1}{4}\nu_-^2
D''\right),\label{type1mismatch}
\end{equation}
where $\nu_-=\nu_i-\nu_s$ and subscripts $p$, $i$, and $s$ refer
to the pump, idler, and the signal, respectively. $D_+ =
1/u_o(\Omega_p/2) - 1/u_p(\Omega_p)$ where $u_o$ ($u_p$) are the
group velocities of the $o$-polarized photon (the pump photon)
inside the crystal and $D^{''}=d^2 K_o/d \Omega^2
|_{\Omega=\Omega_p/2}$ where the wavevector $K_o=\Omega
n_o(\Omega)/c$. $n_o(\Omega)$ is the index of refraction of the
crystal for a given frequency $\Omega$. Note that $D_+$ is
defined differently from the type-II SPDC case.

Therefore, the biphoton amplitude $A(t_+,t_-)$ for type-I SPDC
now becomes \cite{keller}
\begin{eqnarray}
A(t_+,t_-)&=&e^{-i\Omega_p t_+} \int_{-\infty}^\infty
d\nu_p\int_{-\infty}^\infty d\nu_-\int_0^{L}dz\nonumber\\
&\times&e^{-[\nu_p/\sigma_p]^2} e^{-i\left[\nu_p D_+ +
\nu_-^2 D''/4\right]z} \nonumber \\ &\times& e^{-i\nu_p t_+}e^{-i[\nu_-/2]t_-}, \nonumber\\
&\equiv&e^{-i\Omega_pt_+}\Pi(t_+,t_-),\label{type1biphoton}
\end{eqnarray}
where
\begin{equation}
\Pi(t_+,t_-) = \int_0^L dz \frac{1}{\sqrt{z}} e^{-\sigma_p^2[t_+ + D_+ z]^2/4}
e^{it_-^2/[4D^{''}z]}.\label{type1pi}
\end{equation}
Unlike the pulse pumped type-II SPDC, the $\Pi(t_+,t_-)$ function
is symmetric in $t_-$. To simplify the calculation, no spectral
filters are assumed.

\begin{figure}[tbp]
\centerline{\epsfxsize=3.2in\epsffile{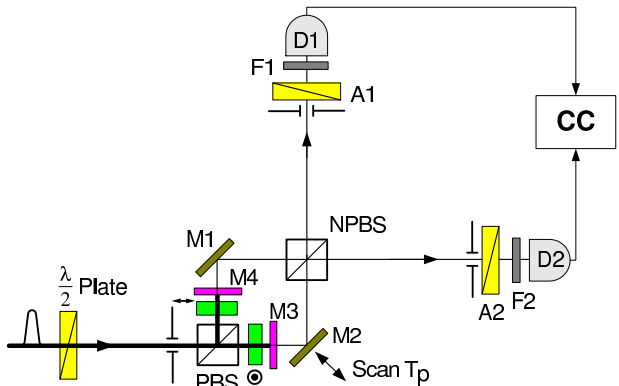}} \caption{Experimental scheme for
two type-I crystal. Now, a type-I BBO crystal is placed in each arm of the MZI and the
optic axes are orthogonally oriented. The output beam splitter (NPBS) of the MZI is now
50/50 non-polarizing beam splitter. }\label{type1setup}
\end{figure}

Having calculated the effective biphoton wave-function of type-I
SPDC, let us now consider the experimental scheme shown in
Fig.\ref{type1setup} and calculate the coincidence counting rate
in detail. Two interfering amplitudes are created from the two
crystals similar to Fig.\ref{type2amplitudes} except that the
biphoton amplitudes now look different in type-I SPDC.  In the
single-mode approximation, the quantum state prepared in the
experimental setup of Fig.\ref{type1setup} is given by
\begin{equation}
|\psi\rangle=|V_1,V_2\rangle + e^{i\varphi}|H_1,H_2\rangle,
\label{MZItypeI}
\end{equation}
where $\varphi$ is the relative phase similar to
Eq.(\ref{MZItypeI}). When the phase $\varphi$ is correctly
chosen, the Bell states $|\Phi^{\pm}\rangle$ can be prepared.
(Note also that by inserting a $\lambda/2$ plate in one output
port of the NPBS, the other two Bell states $|\Psi^\pm\rangle$
can also be prepared.) As we shall show below, there is no need
for any spectral post-selection in this case, however, amplitude
post-selection is assumed because there are possibilities that
the signal and the idler exit at the same output port of the
beamsplitter. This event, however, is not detected since we only
consider the coincidence contributing events. Such amplitude
post-selection is not desirable in principle. Luckily, there is a
way to get around this problem which we shall briefly discuss in
section \ref{type1exp}.

Let us now calculate the coincidence counting rates for the
experimental setup shown in Fig.\ref{type1setup} using the
biphoton amplitude calculated in Eq.(\ref{type1biphoton}). By
using the right-hand coordinate system as in type-II SPDC case,
the coincidence counting rate is given by
\begin{eqnarray}
R_c&\propto&\int dt_+ dt_- |-\sin\theta_1\sin\theta_2\Pi(t_+,t_-) +\nonumber \\&&
\hspace*{0.7in} \cos\theta_1\cos\theta_2e^{-i\Omega_p T_p}\Pi(t_+ + T_p,t_-)
|^2 \nonumber \\
&=&\sin^2\theta_1\sin^2\theta_2 + \cos^2\theta_1\cos^2\theta_2 -\nonumber \\&& 2
G(T_p)\cos(\Omega_p T_p) \sin\theta_1\sin\theta_2\cos\theta_1\cos\theta_2,
\end{eqnarray}
where $G(t)=g(t)/g(0)$ with $$g(t)=\int dt_+ dt_- \Pi(t_+,t_-) \Pi^*(t_+ + t, t_-).$$
The $G(t)$ function gives the envelope of the quantum interference pattern as a function
of $t$.

Therefore, the space-time interference at
$\theta_1=\theta_2=45^\circ$ will show
\begin{equation}
R_c=1-G(T_p)\cos(\Omega_p T_p),\label{type1spacetime}
\end{equation}
and the polarization interference will show
\begin{equation}
R_c=\cos^2(\theta_1+\theta_2) \hspace*{0.5cm} {\rm for}
\hspace*{0.5cm} \Omega_pT_p=0.
\end{equation}

The envelope function of type-I SPDC, $G(T_p)$, differs a lot from
that of the type-II SPDC, $V(T_p)$. $g(T_p)$ is calculated to be

\begin{eqnarray}
g(T_p)&=&\int_{-\infty}^{\infty}dt_+ \int_{-\infty}^{\infty}dt_-
\Pi(t_+,t_-) \Pi^*(t_+ + T_p, t_-) \nonumber \\
&=&\int_0^1du_1 \int_0^1du_2 \int_{-\infty}^{\infty}dt_+
\int_{-\infty}^{\infty}dt_- \nonumber\\
&\times& e^{-\sigma_p^2[t_+ + D_+Lu_1^2]^2/4}e^{it_-^2/[4D''Lu_1^2]} \nonumber \\
&\times& e^{-\sigma_p^2[t_+
+T_p+D_+Lu_2^2]^2/4} e^{-it_-^2/[4D''Lu_2^2]}\nonumber\\
&=&C\int_0^1du_1 \int_0^1du_2 \frac{u_1 u_2}{\sqrt{|u_1^2-u_2^2|}} \nonumber \\ &\times&
e^{-\sigma_p^2\{D_+L[u_1^2-u_2^2]-T_p\}^2/8},\label{type1env}
\end{eqnarray}
where $C$ is a constant and the change of variable, $z_i=u_i^2 L$
($i=1,2$), has been made.

\begin{figure}[tbp]
\centerline{\epsfxsize=3.2in\epsffile{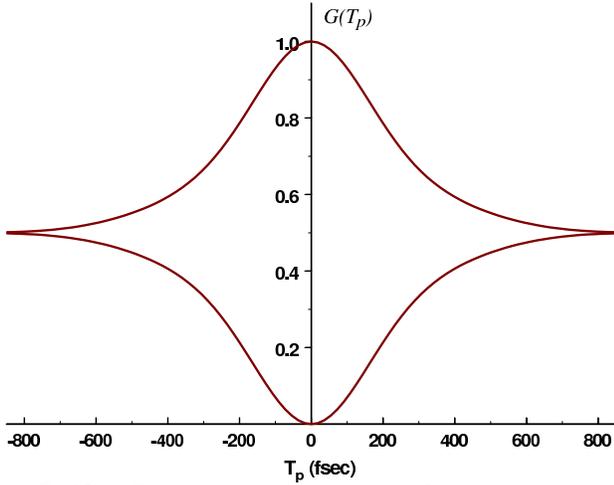}} \caption{ Calculated
envelope of the interference pattern $G(T_p)$ from Eq.(\ref{type1env}) for the
experimental parameters in our experiment. Note that there are long tails of the
interference envelope. }\label{type1envelope}
\end{figure}

It is interesting to find that the envelope function $G(T_p)$ \emph{does not} explicitly
contain the second-order expansion of the phase mismatch term $D''$  at all. This result
is quite surprising since the presence of $D''$ is the principal difference between
type-II SPDC and degenerate type-I SPDC. (This is due to the fact that $D=0$ for
degenerate type-I SPDC.) Note also that the envelope of the space-time interference is
not simply that of the convolution of the pump pulse as in type-II SPDC shown in
Eq.(\ref{spacetime}): it is a complicated function of $D_+ L$ and $\sigma_p$.
Fig.\ref{type1envelope} shows $G(T_p)$ when realistic experimental parameters are
substituted in Eq.(\ref{type1env}).

\section{Bell state preparation using type-I SPDC: Experiment}\label{type1exp}

In the experiment, we use two pieces of type-I BBO crystal cut for
collinear degenerate SPDC. The thickness of the crystals is 3.4
mm each. The pump pulse central wavelength is 400nm and the
average power of the pump beam in each arm of the MZI is
approximately 10mW as in the type-II experiment. The repetition
rate of the pump pulse is approximately 82MHz.

\begin{figure}[tbp]
\centerline{\epsfxsize=3.2in\epsffile{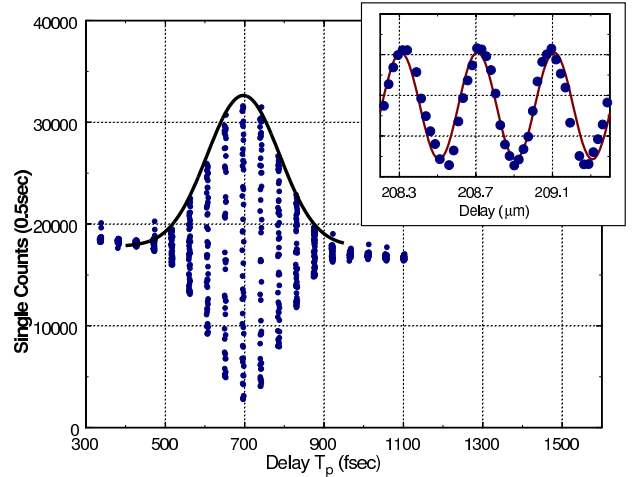}} \caption{Pump pulse
interference. The measured FWHM envelope is approximately 200fsec. Solid line is the
Gaussian fit to the envelope. Each column of data represents the modulation of several
wavelengths with modulation period of 400nm as shown in the inset. The inset has the
same vertical scale as the main figure and it shows the detailed modulation around $T_p
\approx$ 696 fsec. }\label{type1pumpenv}
\end{figure}

We first measured the pump pulse envelope. The BBO crystals are not removed from the MZI
for the pump pulse envelope measurement. This data is shown in Fig.\ref{type1pumpenv}.
Gaussian fitting of the data gives the FWHM equal to 200 fsec. This is to be compared
with the envelope of the quantum interference pattern measured in coincidence counting
rate between the two detectors $D_1$ and $D_2$.

\begin{figure}[tbp]
\centerline{\epsfxsize=3.2in\epsffile{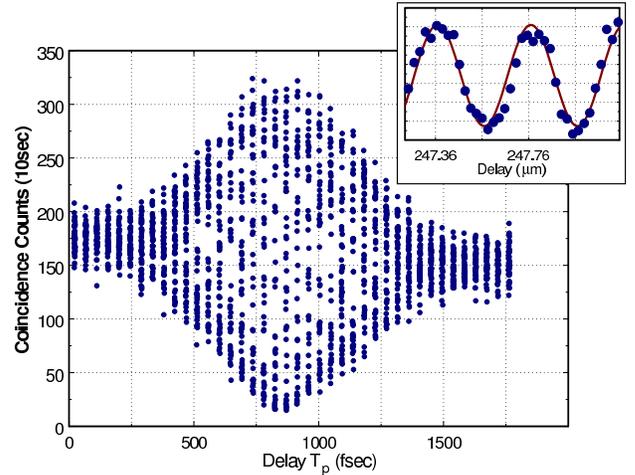}} \caption{Space-time interference
observed in coincidence between the two detectors $D_1$ and $D_2$ for the analyzer
setting $\theta_1=\theta_2=45^\circ$. 10nm filters are used for this measurement. Note
that the FWHM is much bigger than the pump pulse interference shown in
Fig.\ref{type1pumpenv}. 400nm modulation is observed as predicted in
Eq.(\ref{type1spacetime}). The inset has the same vertical scale as the main figure and
it shows the detailed modulation around $T_p \approx$ 826 fsec.}\label{MZItype1}
\end{figure}

To observe the quantum interference, we first block all the
residual pump radiation using additional absorption filters.
Analyzer angles are set at $\theta_1=\theta_2=45^\circ$. The
interference filters used in this measurement have 10nm
bandwidth. As expected, high-visibility quantum interference is
observed, see Fig.\ref{MZItype1}. The FWHM of the interference
envelope is much bigger than that of the pump pulse itself, see
Fig.\ref{type1pumpenv}. Unfortunately, due to rather large
fluctuations in the data, long tails of the interference envelope
predicted by Eq.(\ref{type1env}), see Fig.\ref{type1envelope} are
washed out.

With 40nm filters, the shape of the envelope remained almost the
same, while the width of the envelope and the visibility is
slightly reduced. The reduction of the visibility with 40nm
filters is mainly due to the difficulty in aligning both crystals
with broadband filters.

There are two problems with this method: (i) amplitude
post-selection is assumed, and (ii) the MZI cannot be made very
stable for a long term. In a recently published paper
\cite{kim2}, the second problem was solved by using the
``collinear interferometer" method in which two type-I crystals
are placed collinearly with the optic axes orthogonally oriented.
The first problem, amplitude post-selection, was also removed by
employing \emph{non-degenerate} type-I SPDC in the collinear
scheme \cite{kim3}. Also with collinear method, aligning the
crystals is much easier and the visibility as high as 92\% was
easily obtained with 40nm filters \cite{kim2}. Although the
collinear method and the MZI method look different, the
theoretical description we have presented in the previous
sections applies with no modifications. The theory described in
section \ref{type1theory} explains all the experimental results of
Ref.\cite{kim2} in detail and the experimental results shown in
Ref.\cite{kim3} can be explained by using the theory described in
section \ref{type2theory} by exchanging the polarization label
with frequency label.

\section{Conclusion}\label{conclusion}

We have shown that high-visibility quantum interference can be
observed without using narrow-band filters for both type-II and
type-I SPDC pumped by femtosecond laser pulses. In these methods,
a Mach-Zehnder interferometer is used to coherently add two
biphoton amplitudes from two different nonlinear crystals pumped
by coherent laser pulses. Using this method, maximally entangled
two-photon polarization states, or Bell states, can be
successfully prepared.

It is important to note that biphoton or two-photon amplitudes
generated from coherent pump laser {\em remain coherent} even
though they may originate from different spatial \cite{origin2} or
temporal \cite{kim1,kim5,kim4} domains . As long as the
distinguishing information present in the interfering amplitudes
are erased, high-visibility quantum interference should be
observed.

There is, however, one problem with the method using a
Mach-Zehnder interferometer: keeping the phase coherence is
difficult. This is a rather serious problem especially when one
is interested in using such a method to prepare a Bell state and
use it as a source. The collinear two-photon interferometer
 solves this problem. In this method, two type-I crystals
are placed collinearly in the pump beam path \cite{kim2,kim3}.
Although different geometry is used, the theory presented in
section \ref{type1theory} applies equally for both Mach-Zehnder
interferometer and collinear case.

For the two-crystal scheme using pulse pumped type-II SPDC, the
envelope of the space-time interference pattern is determined
only by the bandwidth of the pump pulse $\sigma_p$, as shown in
section \ref{type2theory} and section \ref{type2exp}. Crystal
parameters do not affect the envelope of the interference pattern
at all. On the other hand, for the two-crystal scheme using pulse
pumped type-I SPDC, the envelope of the interference pattern
strongly depends on the crystal parameters, especially on $D_+ =
1/u_o(\Omega_p/2) - 1/u_p(\Omega_p)$ and the crystal thickness
$L$ as well as the pump bandwidth $\sigma_p$, as shown in section
\ref{type1theory} and section \ref{type1exp}. It is also
interesting to note that the envelope of the interference pattern
does not have explicit dependence on $D''$.

It is important to note the following. To observe the space-time
interference, one can introduce the delay in two ways: (i) in
$t_-$ or (ii) in $t_+$. In the single-crystal SPDC scheme, quantum
interference is observed by introducing the delay $\tau$ in
$t_-$. But in two-crystal SPDC scheme, one can introduce either
in $t_-$ or in $t_+$. In this paper, we have demonstrated
high-visibility quantum interference in femtosecond pulse pumped
SPDC by introducing a delay $T_p$ in $t_+$.

In conclusion, we have demonstrated Bell states preparation schemes using femtosecond
pulse pumped SPDC. In type-II SPDC, the envelope of the interference pattern is exactly
equal to the envelope of the pump pulse convolution. On the other hand, the envelope of
interference pattern from type-I SPDC is much broader than that of the pump
interference. This may be useful if one needs to use femtosecond pulse pumped SPDC, yet
requires that two-photon amplitudes are distributed in time more than the pump pulse
itself. Type-I SPDC has an advantage over type-II SPDC: two crystals can be easily used
collinearly. As demonstrated in \cite{kim2,kim3}, such a method will serve as a good
source of entangled photon pairs for experiments which require accurate timing to
overlap biphotons from different domains.

\section*{Acknowledgement}
We would like to thank V. Berardi and L.-A. Wu for their help
during the last part of the experiment. This research was
supported, in part, by the Office of Naval Research, ARDA and the
National Security Agency.

This paper is dedicated to the memory of our colleague and teacher
D.N. Klyshko.


\vspace*{-2mm}

\begin{references}

\bibitem{Schrodinger} E. Schr\"{o}dinger, Naturwissenschaften
\textbf{23}, 807 (1935); \textbf{23}, 823 (1935); \textbf{23},
844 (1935); the English translation appears in \emph{Quantum
Theory and Measurement}, edited by J.A. Wheeler and W.H. Zurek
(Princeton University Press, New York, 1983).

\bibitem{EPR}  A. Einstein, B. Podolsky, and N. Rosen, Phys. Rev. \textbf{47},
777 (1935).

\bibitem{Bohm} D. Bohm, \emph{Quantum Theory} (Prentice-Hall Inc.,
New York, 1951).

\bibitem{Bell} J.S. Bell, \emph{Speakable and unspeakable in
quantum mechanics}, (Cambridge University Press, New York, 1987).

\bibitem{clauser} S.J. Freedman and J.F. Clauser, Phys. Rev.
Lett. \textbf{28}, 938 (1972); J.F. Clauser and A. Shimony, Rep.
Prog. Phys. \textbf{41}, 1881 (1978).

\bibitem{aspect} A. Aspect, P. Grangier, and G.
Roger, Phys. Rev. Lett. \textbf{47}, 460 (1981); A. Aspect, J.
Dalibard, and G. Roger, Phys. Rev. Lett. \textbf{49}, 1804
(1982); A. Aspect, P. Grangier, G. Roger, Phys. Rev. Lett.
\textbf{49}, 91 (1982).

\bibitem{shihalley} C.O. Alley and Y.H. Shih, {\em Proc. of 2nd Int. Symp.
Foundations of Quantum Mechanics}, ed. M. Namiki (Physical
Society of Japan, Tokyo) (1987); Y.H. Shih and C.O. Alley, Phys.
Rev. Lett. {\bf 61}, 2921 (1988).

\bibitem{usingspdc} J.G. Rarity and P.R. Tapster, Phys. Rev. Lett.
\textbf{64}, 2495 (1990); G. Weihs, T. Jennewein, C. Simon, H.
Weinfurter, and A. Zeilinger, Phys. Rev. Lett. \textbf{81}, 5039
(1998).

\bibitem{crypto} T. Jennewein, C. Simon, G. Weihs, H.
Weinfurter, and A. Zeilinger, Phys. Rev. Lett. \textbf{84}, 4729
(2000); D.S. Naik, C.G. Peterson, A.G. White, A.J. Berglund, and
P.G. Kwiat, Phys. Rev. Lett. \textbf{84}, 4733 (2000); W. Tittel,
J. Brendel, H. Zbinden, and N. Gisin, Phys. Rev. Lett.
\textbf{84}, 4737 (2000).

\bibitem{teleportation} C.H. Bennett, G. Brassard, C. Cr\'{e}peau,
R. Jozsa,A. Peres, and W.K. Wootters , Phys. Rev. Lett. \textbf{70}, 1895 (1993); D.
Bouwmeester, J.-W. Pan, M. Eibl, H. Weinfurter, and A. Zeilinger , Nature \textbf{390},
575 (1997); D. Boschi, S. Branca, F. De Martini, L. Hardy, and S. Popescu, Phys. Rev.
Lett. \textbf{80}, 1121 (1998); A. Furusawa, J.L. S$\o$rensen, S.L. Braunstein, C.A.
Fuchs, H.J. Kimble, E.S. Polzik, Science \textbf{282}, 706 (1998); Y.-H. Kim, S.P. Kulik,
and Y.H. Shih, Phys. Rev. Lett. \textbf{86}, 1370 (2001).

\bibitem{SPDC} D.N. Klyshko, {\em Photons and Nonlinear Optics},
(Gordon and Breach, 1988).

\bibitem{klyshkoSPDC} D.N. Klyshko, Soviet Phys-JETP Lett. \textbf{6}, 23 (1967).

\bibitem{zeldovich} Ya. B. Zel'dovich and D.N. Klyshko, JETP Lett. \textbf{9}, 40 (1969).

\bibitem{burnham} D.C. Burnham and D.L. Weinberg, Phys. Rev. Lett. \textbf{25}, 84 (1970).

\bibitem{klyshko2} D.N. Klyshko, Sov. J. Quantum Electron. \textbf{7}, 591
(1977); M.F. Vlasenko, G. Kh. Kitaeva, and A.N. Penin, Sov. J.
Quantum Electron. \textbf{10}, 252 (1980).

\bibitem{calibration} D.N. Klyshko, Sov. J. Quantum Electron. \textbf{10}, 1112 (1980);
A.A. Malygin, A.N. Penin, and A.V. Sergienko, JETP Lett.
\textbf{33}, 477 (1981); J.G. Rarity, K.D. Ridley, and P.R.
Tapster, Applied Optics \textbf{26}, 4616 (1987).

\bibitem{HOM} C.K. Hong, Z.Y. Ou, and L. Mandel, Phys. Rev. Lett. \textbf{59}, 2044 (1987).

\bibitem{shihsergienko} Y.H. Shih and A.V. Sergienko, Phys. Lett. A \textbf{186}, 29 (1994);
Phys. Lett. A \textbf{191}, 201 (1994).

\bibitem{garuccio} L. De Caro and A. Garuccio, Phys. Rev. A {\bf
50}, R2803 (1994).

\bibitem{kwiat1} P.G. Kwiat, K. Mattle, H. Weinfurter, A. Zeilinger,
A.V. Sergienko, and Y.H. Shih, Phys. Rev. Lett. {\bf 75}, 4337 (1995);

\bibitem{kwiat2} P.G. Kwiat, E. Waks, A.G. White, I. Appelbaum, and
P.H. Eberhard, Phys. Rev. A {\bf 60}, R773 (1999).

\bibitem{Burlakov} A.V. Burlakov, M.V. Chekhova, O.A. Karabutova, D.N. Klyshko,
and S.P. Kulik, Phys. Rev. A \textbf{60}, R4209 (1999); A.V.
Burlakov and D.N. Klyshko, JETP Lett. \textbf{69}, 839 (1999).

\bibitem{keller}T.E. Keller and M.H. Rubin, Phys. Rev. A {\bf 56},
1534 (1997).

\bibitem{sergienko} A.V. Sergienko, M. Atat\"{u}re,
Z. Walton, G. Jaeger, B.E.A. Saleh, and M.C. Teich, Phys. Rev. A
\textbf{60}, R2622 (1999).

\bibitem{femto} W.P. Grice and I.A. Walmsley, Phys. Rev. A {\bf 56},
1627 (1997); G. Di Giuseppe, L. Haiberger, F. De Martini, A.V. Sergienko, Phys. Rev. A
{\bf 56}, R21 (1997); W.P. Grice, R. Erdmann, I.A. Walmsley, D. Branning, Phys. Rev. A
{\bf 57}, R2289 (1998).

\bibitem{branning} D. Branning, W.P. Grice, R. Erdmann, and I.A. Walmsley,
Phys. Rev. Lett \textbf{83}, 955 (1999); Phys. Rev. A
\textbf{62}, 013814 (2000).

\bibitem{notebranning} It should be noted that Branning \emph{et al} do not
claim Bell state preparation in Ref.\cite{branning}. In this paper, we simply point out
that the experimental scheme demonstrated in Ref.\cite{branning} cannot be used to
prepare a Bell state which is of our interest.

\bibitem{atature} M. Atat\"{u}re, A.V. Sergienko, B.E.A. Saleh, and M.C.
Teich, Phys. Rev. Lett. \textbf{84}, 618 (2000).

\bibitem{comment}  Y.-H. Kim, S.P. Kulik, M.H. Rubin, and Y.H. Shih, \verb!quant-ph/0006003!.

\bibitem{Rubin} M.H. Rubin, D.N. Klyshko, Y.H. Shih, and A.V. Sergienko
, Phys. Rev. A {\bf 50}, 5122 (1994).

\bibitem{Glauber} R.J. Glauber, Phys. Rev. {\bf 130}, 2529
(1963).

\bibitem{kim1} Y.-H. Kim, V. Berardi, M.V. Chekhova, A. Garuccio,
and Y.H. Shih, Phys. Rev. A \textbf{62}, 043820 (2000).

\bibitem{origin} The presence of the first-order interference in Branning
\emph{et al}'s scheme is not surprising. Similar effects are already observed by many
authors, see Ref.\cite{kim1,origin2,kim5}. First-order interference, although very
interesting, is not desirable here and there should not be any first-order interference
the purpose of Bell state preparation.

\bibitem{notebranning2} Branning \emph{et al}'s scheme uses a $\lambda/2$ plate and a polarizing
beam splitter. This setup is essentially equivalent to the scheme with polarizers set at
$45^\circ$.  As pointed out in Ref.\cite{notebranning}, Branning \emph{et al} do not
claim Bell state preparation and therefore there is no need for varying the angles of the
analyzers. We are, however, focused on Bell state preparation. Therefore, it is
necessary that angles of the analyzers can be varied arbitrarily.

\bibitem{origin2} X.Y. Zou, L.J. Wang, and L. Mandel, Phys. Rev.
Lett. \textbf{67}, 318 (1991); T.J. Herzog, J.G. Rarity, H.
Weinfurter, and A. Zeilinger, Phys. Rev. Lett. \textbf{72}, 629
(1994); A.V. Burlakov, M.V. Chekhova, D.N. Klyshko, S.P. Kulik,
A.N. Penin, Y.H. Shih, and D.V. Strekalov, Phys. Rev. A
\textbf{56}, 3214 (1997).

\bibitem{kim5} Y.-H. Kim, M.V. Chekhova, S.P. Kulik, Y.H. Shih,
and M.H. Rubin, Phys. Rev. A \textbf{61}, 051803(R) (2000).

\bibitem{similar} Kwiat \emph{et al} proposed a similar scheme with cw pumped SPDC for
the purpose of loophole-free test of Bell's inequality, see
Ref.\cite{kwiat3}.

\bibitem{kwiat3} P.G. Kwiat, P.H. Eberhard, A.M. Steinberg, and
R.Y. Chiao, Phys. Rev. A \textbf{49}, 3209 (1994).

\bibitem{walkoff} For 3.4mm type-II phased matched (400/800nm) BBO
crystal, the maximum transverse walk-off between the e-ray and the
o-ray is $250\mu$m. Compared to 2.5mm beam diameter, this is not
big, but certainly not negligible.

\bibitem{kim2} Y.-H. Kim, S.P. Kulik, and Y.H. Shih, Phys. Rev. A
\textbf{62}, 011802(R) (2000).

\bibitem{kim3} Y.-H. Kim, S.P. Kulik, and Y.H. Shih,
\verb!quant-ph/0007067!.

\bibitem{kim4} Y.-H. Kim, M.V. Chekhova, S.P. Kulik, and Y.H.
Shih, Phys. Rev. A \textbf{60}, R37 (1999).
\end{references}
\end{document}